\documentstyle[aps,prb,multicol,graphicx]{revtex}

\begin{document}
\draft

\author{L. Ghivelder\cite{corresp} and I. Abrego Castillo}
\address{Instituto de F\'{\i }sica, Universidade Federal do Rio de Janeiro, C.P.\\
68528, Rio de Janeiro, RJ 21945-970, Brazil}
\author{M. A. Gusm\~{a}o}
\address{Instituto de F\'{\i }sica, Universidade Federal do Rio Grande do Sul,\\
C.P.15051, Porto Alegre, RS 91501-970, Brazil}
\author{J. A. Alonso}
\address{Instituto de Ciencia de Materiales de Madrid, C.S.I.C., Cantoblanco, 28049\\
Madrid, Spain}
\author{L. F. Cohen}
\address{Blackett Laboratory, Imperial College, London SW7 2BZ, United Kingdom}
\title{Specific heat and magnetic order in LaMnO$_{3+\delta}$}
\maketitle

\begin{abstract}
Magnetic and specific-heat measurements are performed in three different
samples of LaMnO$_{3+\delta }$, with $\delta =0.11$, 0.15 and 0.26,
presenting important disorder effects, such as carrier localization, due to
high amounts of La and Mn vacancies. For the samples with $\delta =0.11$ and
0.15, magnetic measurements show signatures of a two-step transition: as the
temperature is lowered, the system enters a ferromagnetic phase followed by
a disorder-induced cluster-glass state. Spin-wave-like contributions and an
unexpected large linear term are observed in the specific heat as a function
of temperature. In the sample with the highest vacancy content, $\delta
=0.26 $, the disorder is sufficient to suppress even short-range
ferromagnetic order and yield a spin-glass-like state.
\end{abstract}


\begin{multicols}{2}

\section{Introduction}

Hole-doped perovskite-type manganese oxides have attracted considerable
interest in recent years, motivated by the observation of colossal
magnetoresistance (CMR) in numerous related compounds, and the great variety
of magnetic and transport properties in this class of materials.\cite{Review}
Among the La-based systems, the ground state of the stoichiometric parent
compound LaMnO$_{3}$ is insulating A-type antiferromagnetic (AF), which is
attributed to a cooperative effect of orbital ordering and superexchange
interactions.\cite{SuExch} Substitution of a fraction $x$ of La$^{3+}$ by
divalent cations such as Sr$^{2+}$, Ca$^{2+}$ or Ba$^{2+}$ causes the
conversion of a proportional number of Mn$^{3+}$ to Mn$^{4+}$. At certain
doping ranges ($0.2 \lesssim x \lesssim 0.5$) this induces a metal-insulator
transition and the appearance of a ferromagnetic (FM) state. The
simultaneous FM and metallic transitions have been qualitatively explained
by the double-exchange (DE) model,\cite{Zen} which considers the magnetic
coupling between Mn$^{3+}$ and Mn$^{4+}$ resulting from the motion of an
electron between the two partially filled $d$ shells. Nevertheless, this DE
mechanism does not account for several experimental results, and it has been
claimed\cite{Millis} that a Jahn-Teller type electron-phonon coupling plays
an important role in explaining the large magnetoresistive effects.

Conversion of Mn$^{3+}$ to Mn$^{4+}$ can also be achieved by the presence of
non-stoichiometric oxygen in undoped LaMnO$_{3+\delta}$, with a nominal Mn$
^{4+}$ content of $2\delta$. For simplicity this is the crystallographic
representation used in the present work and in most other studies in this
system. However, it does not reflect the fact that the system contains
randomly distributed La and Mn vacancies rather than oxygen excess, which
can not be accommodated interstitially in the lattice.\cite{Oxy} The actual
crystallographic formula is better written as La$_{1-x}$Mn$_{1-y}$O$_3$. By
varying the oxygen stoichiometry the resulting compounds display a wide
variety of structural and magnetic phases, previously studied by x-rays and
neutron scattering,\cite{Ritter,Alon2} as well as magnetic and transport
measurements.\cite{Ritter,Les} It is well known that the low-temperature
magnetic phase of non-stoichiometric LaMnO$_{3+\delta}$ changes from AF to
FM for small values of $\delta$ due to the DE interaction caused by the
presence of Mn$^{4+}$ ions in the sample. However, unlike the cation-doped
systems, the material remains insulating at all temperatures, and the FM
transition temperature decreases for increasing content of Mn$^{4+}$. The
relevant fact to be considered appears to be the competing effect between La
vacancies, enhancing the Mn$^{3+}$-Mn$^{4+}$ DE interaction, and Mn
vacancies which introduce considerable disorder in the lattice. For large
values of $\delta$ the FM order is suppressed, and the low-temperature phase
is better described by a spin-glass-like state.\cite{Ritter,Les}

The competing effect between cation and manganese vacancies makes LaMnO$
_{3+\delta }$ a model system for studying magnetic interactions and disorder
effects in mixed-valence manganites. In order to achieve a better
understanding of the low temperature properties of this system we have
performed magnetic and specific-heat measurements in three different samples
of non-stoichiometric LaMnO$_{3+\delta}$. Magnetic data show signatures of a
double transition: as the temperature is lowered, the system first orders
ferromagnetically in small weakly-connected clusters, and then changes to a
cluster-glass phase. Results of low-temperature specific-heat measurements
show an unexpectedly large linear coefficient and a spin-wave contribution.
This is interpreted in terms of the existence of disorder-induced
charge-localization in these compounds.

\section{Experiments}

The bulk samples of LaMnO$_{3+\delta}$ investigated in the present study
were thoroughly characterized in Refs.\ \onlinecite{Alon2,Les,Alon3}. They
were prepared in polycrystalline form by a citrate technique, as described
elsewhere.\cite{Alon2} The products were annealed at 1100 $^\circ$C in air
(Sample 1), 1000 $^\circ$C in air (Sample 2) and 1000 $^\circ$C under 200
bar of O$_2$ (Sample 3). The determination of $\delta$ was initially
performed by thermogravimetric analysis. The final materials were
characterized by x-ray diffraction. Neutron powder diffraction diagrams were
also collected in the temperature range 2-250$\;$K. The Rietveld method was
used to refine the crystal and magnetic structures.

The neutron-diffraction refinements showed that all investigated samples
have stoichiometric oxygen content of $3.00\pm 0.05$. The Mn$^{4+}$ content
was calculated from the vacancy concentration of La and Mn determined from
the neutron data, and found to be in good agreement with the
thermogravimetric analysis. Sample 1, with $\delta =0.11$ and 23\% of Mn$
^{4+}$, consists of a mixture of a main orthorhombic phase (64\%) and a
minor rhombohedral phase (36\%). Sample 2, with $\delta =0.15$ and 33\% of Mn
$^{4+}$, and Sample 3, with $\delta =0.26$ and 52\% of Mn$^{4+}$, both have
rhombohedral symmetry. Samples 1 and 2 showed a FM ordered structure at low
temperatures (with some canting observed in Sample 2), whereas Sample 3
showed spin-glass-like signatures. Transport measurements\cite{Les} revealed
that all the studied compounds are insulating down to low temperatures, with
a typical semiconductor-like behavior. Selected sample parameters are
summarized in Table I.

\bigskip

\vbox{\narrowtext
\begin{table}
\caption{Selected physical parameters and preparation conditions of the LaMnO
$_{3+\delta}$ samples. The actual crystallographic formula is better written
as La$_{1-x}$Mn$_{1-y}$O$_3$. The FM transition temperature is given by $T_c$
.}
\label{tab:1}
\begin{tabular}{c|ccc}
Sample & 1 & 2 & 3 \\ 
$\delta $ & 0.11 & 0.15 & 0.26 \\ 
$x$, $y$ & 0.022, 0.054 & 0.029, 0.069 & 0.029, 0.128 \\ 
Mn$^{4+}$ (\%) & 23 & 33 & 52 \\ 
Prep.\ Conditions & 1100 $^\circ$C/air & 1000 $^\circ$C/air & 1000 
$^\circ$C/O$_{2}$ \\ 
Cryst.\ Structure & Ortho./Rhomb. & Rhomb. & Rhomb. \\ 
$T_c$ (K) & 154 & 142 & --
\end{tabular}
\end{table}
}

\narrowtext
\begin{figure}
\begin{center}
\includegraphics[width=7.50cm,angle=0,clip]{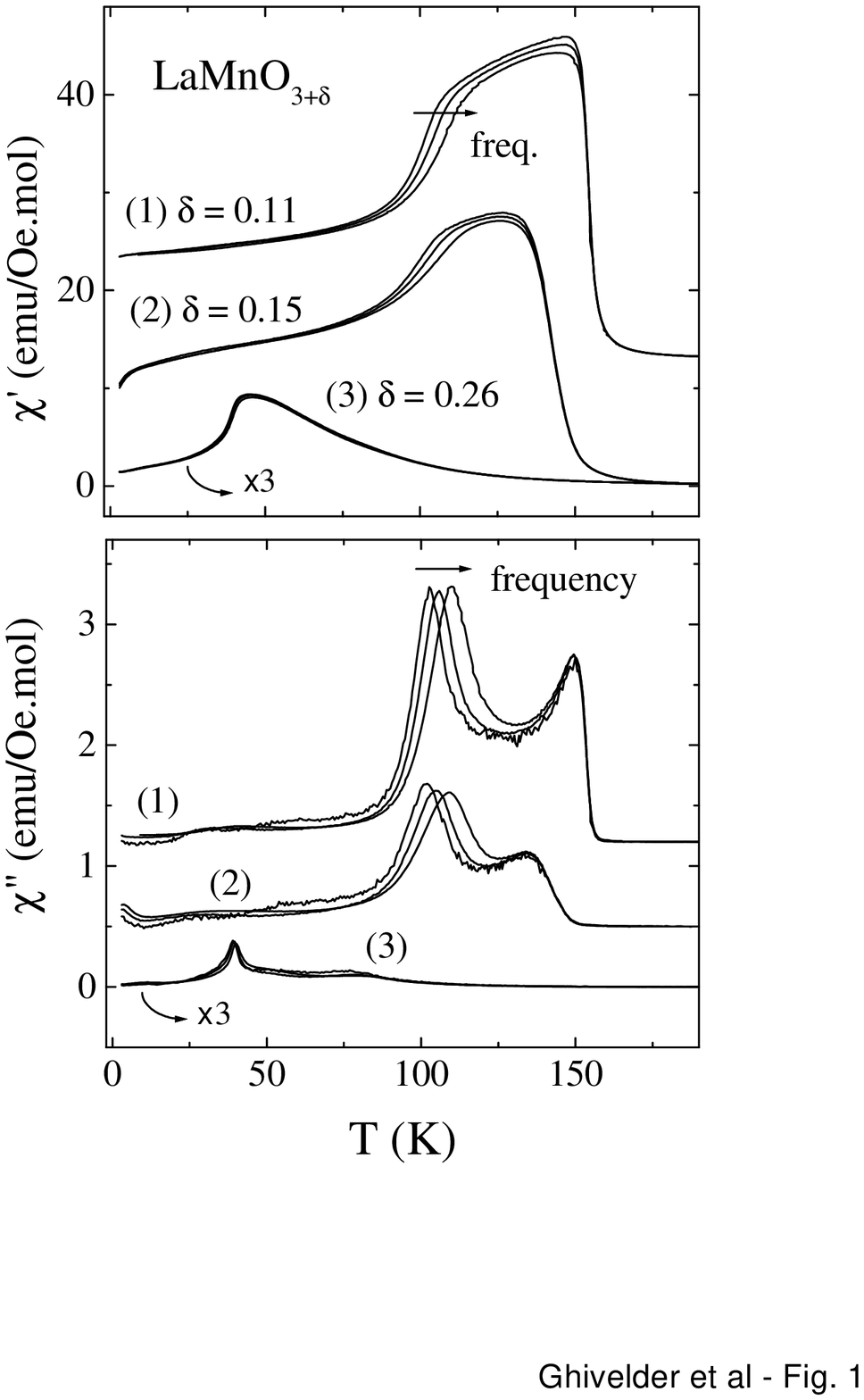}
\end{center}
\caption{Real and imaginary parts of the AC susceptibility of LaMnO$
_{3+\delta }$, measured in an alternating field $h_{ac}$ = 1$\;$Oe and
frequencies $f = 25$, 125 and 1000$\;$Hz. For samples with $\delta = 0.11$
and 0.15 (Samples 1 and 2) the data is shifted vertically for clarity. For
the sample with $\delta = 0.26$ (Sample 3) the results are multiplied by 3.}
\label{fig:1}
\end{figure}

In the present study, specific-heat results were obtained from 4.5 to 200 K
with an automated quasi-adiabatic pulse technique. The absolute accuracy of
the data, checked against a copper sample, is better than 3\%. The measured
samples had masses of approximately 50$\;$mg. Detailed AC susceptibility and
DC magnetization measurements where performed in a commercial magnetometer
(Quantum Design PPMS). The FM transition temperatures of Samples 1 and 2,
obtained from AC susceptibility data, are also shown in Table I.

\section{Magnetic Measurements}

Figure\ 1 shows the AC susceptibility of LaMnO$_{3+\delta }$. Real and
imaginary parts, respectively $\chi ^{\prime }$ and $\chi ^{\prime \prime }$
, were measured in zero DC field, with an alternating field $h_{ac}=1\;$Oe,
and frequencies of 25, 125 and 1000 Hz. Results for Sample 3 are multiplied
by a factor of 3. Part
\end{multicols}
\widetext
\begin{figure}
\begin{center}
\includegraphics[width=16.0cm,angle=0,clip]{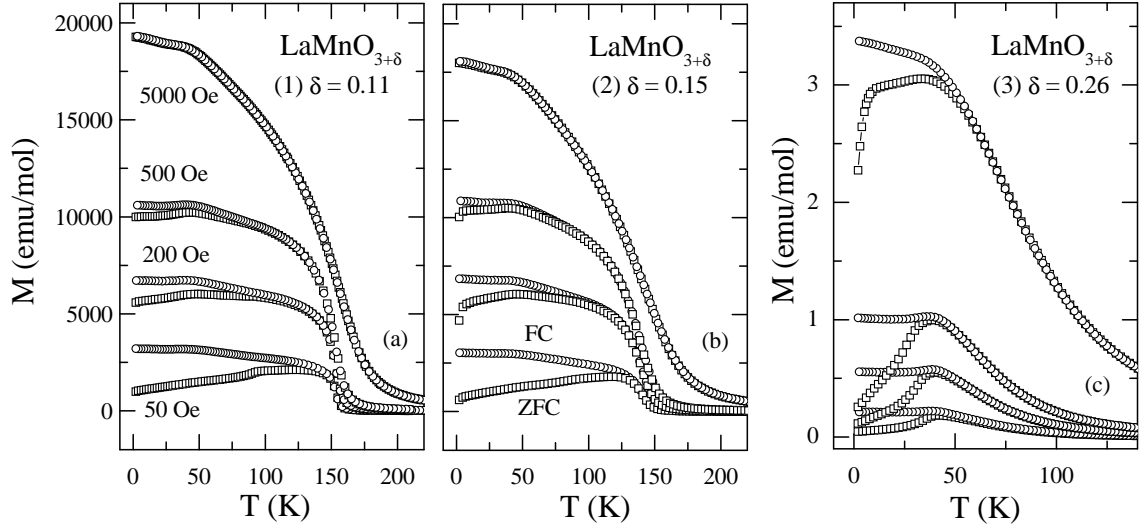}
\end{center}
\caption{Field-cooled (FC) and zero-field-cooled (ZFC) DC magnetization of
LaMnO$_{3+\delta }$. The applied field, from bottom to top in the figures,
is 50, 200, 500 and 5000$\;$Oe.}
\label{fig:2}
\end{figure}

\begin{multicols}{2}
\noindent of the data was shifted vertically for clarity. The
first point to note is a pronounced FM transition, observed at 154 and 142 K
for Samples 1 ($\delta =0.11$) and 2 ($\delta =0.15$), respectively. The
values of T$_{c}$ were determined from the maximum derivative in $\chi
^{\prime }$. For Sample 3, with the higher vacancy content ($\delta =0.26$),
at 48$\;$K we observe a much lower cusp-like anomaly in $\chi ^{\prime }$,
typical of a spin glass behavior. As mentioned in the introduction, the
evolution from FM to spin glass features for increasing oxygen content in
LaMnO$_{3+\delta }$ was previously observed in the literature.\cite
{Ritter,Les}

Moreover, it is most interesting to note in Fig.\ 1 that the results for the
two FM samples show a double-peak structure and a frequency dependence of
the imaginary component, $\chi^{\prime\prime}$. The high-temperature peak is
frequency independent, whereas the position of the low-temperature peak
strongly depends on the measuring frequency. The maximum in $
\chi^{\prime\prime}$ shifts to higher temperatures as the frequency
increases. These are clear signatures of a cluster-glass behavior, as
previously reported for other manganite\cite{doubleX} and cobaltite\cite
{Co1,Co2} systems. The high-temperature peak signals the onset of FM order,
whereas the low-temperature frequency-dependent peak is associated with
freezing of the cluster magnetic moments. In connection with the
low-temperature peak in $\chi^{\prime\prime}$, a frequency-dependent
shoulder can be observed in the real component $\chi^{\prime}$. Results for
Sample 3 also show a distinct frequency dependence in $\chi^{\prime}$, not
visible in the scale of the figure.

In order to probe disorder-induced features in the system, we have measured
the field-cooled (FC) and zero-field-cooled (ZFC) magnetization of the
studied samples. Figures\ 2(a) and 2(b) display the results for Samples 1
and 2 respectively. The low-field data was taken with $H=50\;$Oe. A
pronounced irreversibility is observed, again indicative of a disordered
state. In our results the irreversibility starts just below T$_{c}$, which
is the typical behavior of a cluster-glass phase, whereas in
reentrant-spin-glass systems irreversibility occurs far below T$_{c}$.

As the field increases the irreversible behavior is reduced, and is no
longer present at $H=5000\;$Oe. Measurements of AC susceptibility with an
applied DC field (not shown) confirm that the frequency dependence in $\chi
^{\prime }$ disappears with increasing fields. These results show that the
application of a DC field tends to align the cluster moments, and stabilizes
a reversible FM ordered state. In the magnetization results for Sample 3,
shown in Fig.\ 2(c), the behavior is quite different. The magnetization peak
is more than two orders of magnitude lower than in the other samples, and
the irreversibility persists with higher applied DC field, which confirms
the standard spin-glass features in the high-vacancy sample. The difference
between the ZFC and FC magnetizations is much higher in the cluster-glass
phase (Samples 1 and 2) compared to the spin-glass phase (Sample 3),
reflecting the presence of FM order within the clusters.\cite{Co1}

Isothermal $M$ vs.\ $H$ curves measured at 10$\;$K are plotted in Fig.\ 3.
For the FM samples (1 and 2) the magnetization saturates at fields of the
order of 1--2$\;$T. The saturation values are $3.70\,\mu _{B}$ and $
3.57\,\mu _{B}$ for Samples 1 and 2, respectively. The magnetic moment
expected from the spin contribution is $gS\,\mu _{B}$, where $S$ is the spin
of the ion, which is 3/2 for Mn$^{4+}$ and 2 for Mn$^{3+}$, and the
gyromagnetic factor $g=2$ in both cases. Taking into account the relative
concentrations of Mn$^{4+}$ and Mn$^{3+}$ in the compounds, we get an
effective moment of $3.77\,\mu _{B}$ for Sample 1 (23\% Mn$^{4+}$), and $
3.67\,\mu _{B}$ for Sample 2 (33\% 
\narrowtext
\begin{figure}
\begin{center}
\includegraphics[width=7.50cm,angle=0,clip]{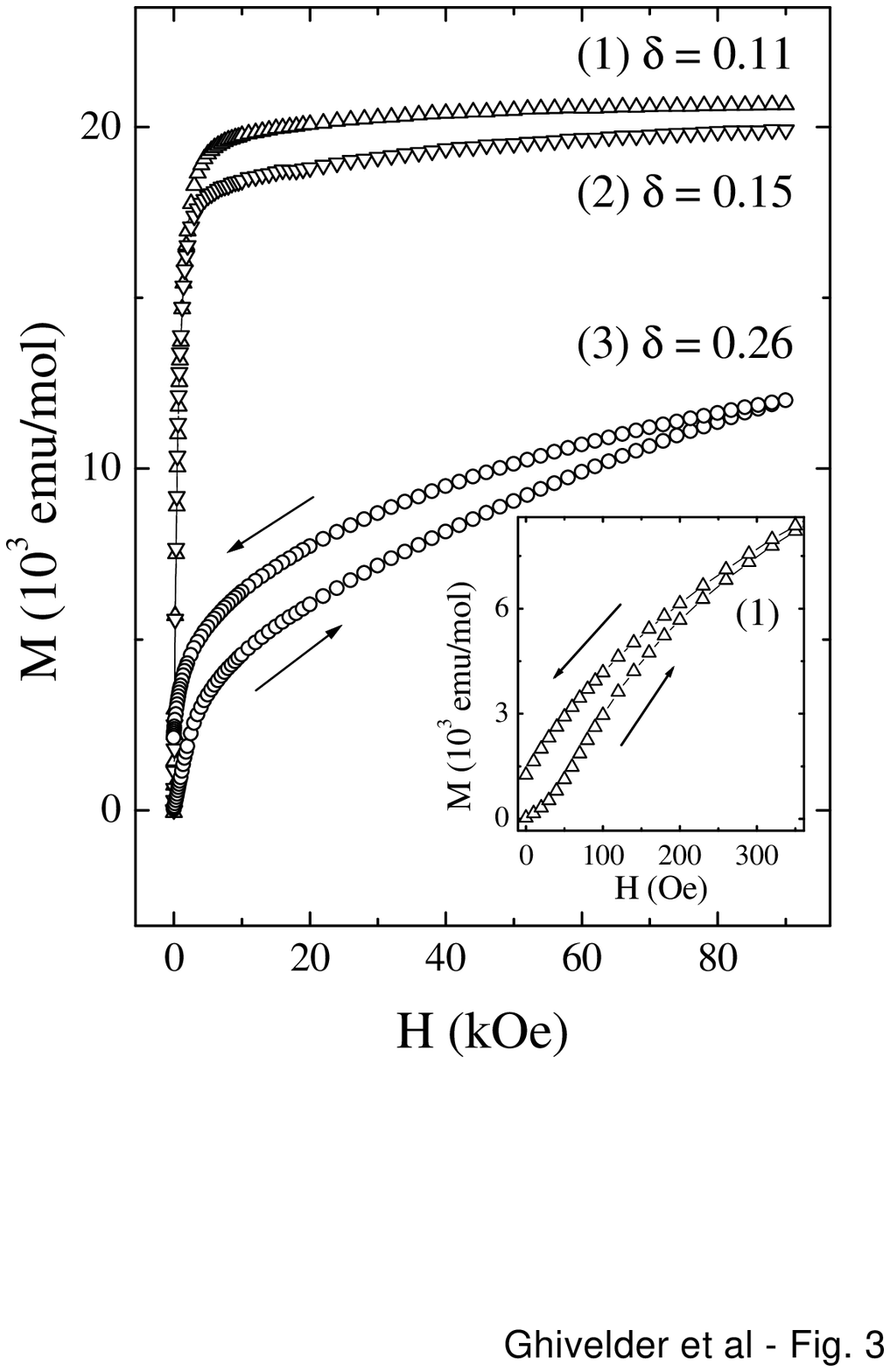}
\end{center}
\caption{Magnetization as a function of field of LaMnO$_{3+\delta }$,
measured at 10$\;$K. The inset shows low-field data for Sample 1. The arrows
indicate measurements increasing and decreasing the field.}
\label{fig:3}
\end{figure}

\noindent Mn$^{4+}$). This prediction virtually
coincides with the values observed experimentally, indicating that the
applied field fully polarizes the FM clusters. Hysteresis is observed at
very low fields, up to about 400 Oe, as shown for Sample 1 in the inset of
Fig.\ 3. This hysteresis is consistent with the $M$ vs.\ $T$ data of Fig.\
2, and is attributed to the low field cluster-glass nature of the samples.
For Sample 3 (52\% Mn$^{4+}$), the low-temperature magnetization does not
saturate at our highest field, and a large hysteretic behavior is observed.
At 9$\;$T the measured magnetic moment is $2.15\,\mu _{B}$, much smaller
than the predicted value of $3.48\,\mu _{B}$. This is an additional
indication of the spin-glass-like properties of this sample.

In order to verify the consistency of our magnetic results, we have
performed the same measurements on another similar series of LaMnO$
_{3+\delta}$ samples. The cluster-glass behavior of the intermediate-vacancy
FM samples, i.e., the frequency-dependent AC susceptibility and the
irreversibility in low-field magnetization, were confirmed to exist in this
second series of samples.

\section{Specific Heat Measurements}

Figure\ 4(a) shows the specific heat of the investigated samples
plotted as $C/T$ vs. $T^{2}$, in the temperature range of
4.5--15$\;$K. For comparison, measurements on
La$_{0.90}$Ca$_{0.10}$MnO$_{3}$, a ferromagnetic insulator, and
\nopagebreak on La$_{0.67}$Ca$_{0.33}$MnO$_{3}$, ~~a ~ferromagnetic
~metal, ~~are ~also 

\narrowtext
\begin{figure}
\begin{center}
\includegraphics[width=7.50cm,angle=0,clip]{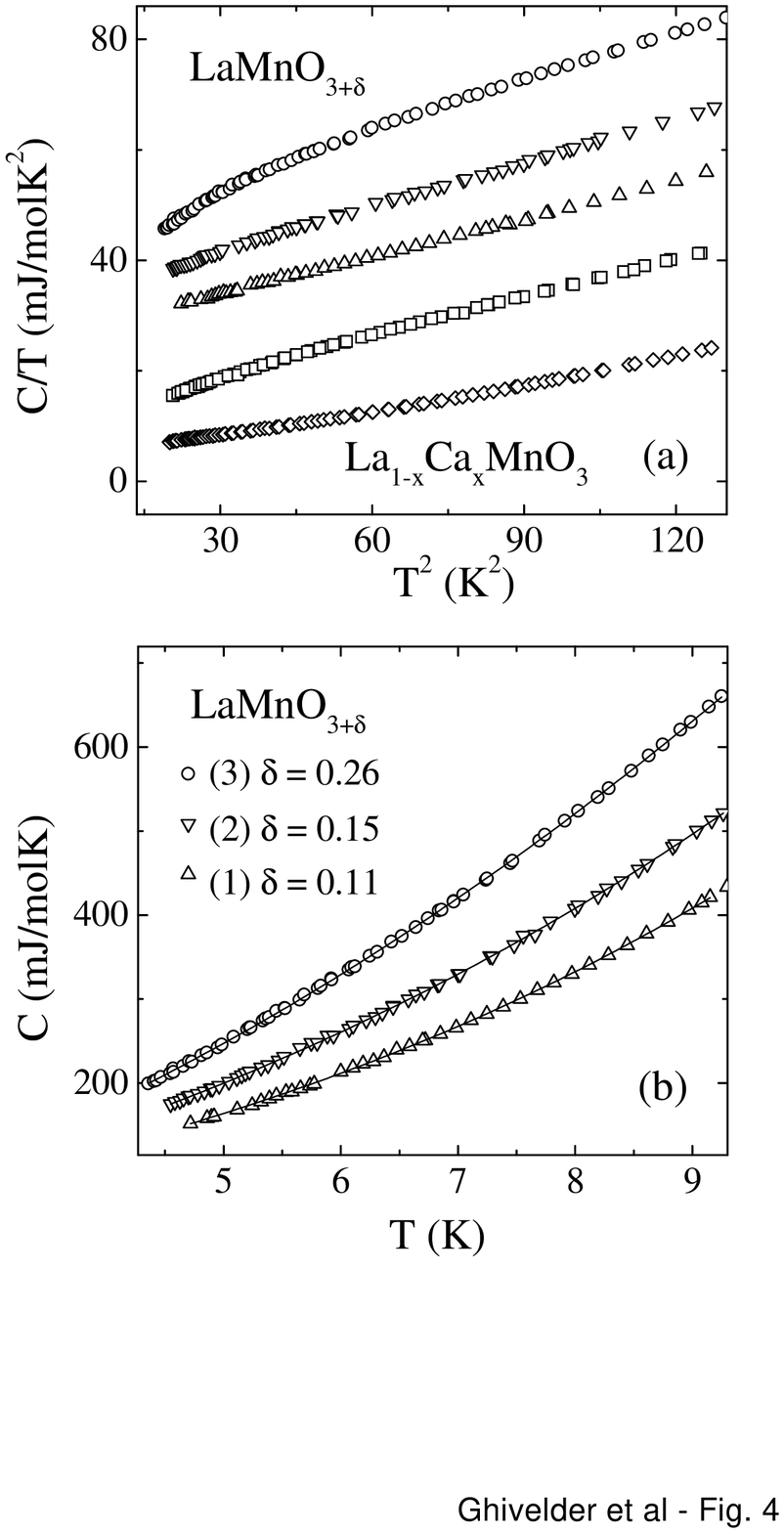}
\end{center}
\caption{(a) From top to bottom, low-temperature specific heat, plotted as $
C/T$ vs.\ $T^{2}$, for LaMnO$_{3+\delta}$ with $\delta = 0.26$ (circles),
0.15 (down triangles), and 0.11 (up triangles), and for La$_{1-x}$Ca$_x$MnO$
_3$ with $x = 0.11$ (squares) and 0.33 (diamonds). (b) Plot of $C$ vs.\ $T$
for LaMnO$_{3+\delta }$; the solid lines are fitted curves, as discussed in
the text.}
\label{fig:4}
\end{figure}

\noindent displayed.\cite{Ghiv} The latter has the same Mn$^{4+}$
content as in LaMnO$_{3+\delta }$ with $\delta =0.15$. However, it
is clear from the figure that the heat capacity is considerably higher
in LaMnO$_{3+\delta }$ as compared to the Ca-doped compounds. In order
to interpret these results and evaluate the different contributions to
the specific heat, the low-temperature data of each studied sample
were fitted to the expression
\begin{equation}
C=\gamma T+\beta T^{3}+BT^{3/2}.
\end{equation}
The linear coefficient $\gamma $ is usually attributed to charge carriers,
and is proportional to the density of states at the Fermi level. However,
transport measurements showed that all the investigated samples of LaMnO$
_{3+\delta }$ are insulating, and the appearance of a linear term in the
specific heat must be more carefully interpreted. The lattice contribution
is given by $\beta T^{3}$. A higher-order lattice term proportional to $
T^{5} $ was not needed to fit the data in the temperature range up to 10$\;$
K. The term $BT^{3/2}$ is associated with FM spin-wave excitations. The
coefficient 
\end{multicols}

\widetext
\begin{table}[tbp]
\caption{Fitting results of the low-temperature specific heat of LaMnO$%
_{3+\delta }$. The linear coefficient is given by $\gamma$. The Debye
temperature $\theta_D$ is obtained from the cubic coefficient $\beta $, and
the spin-wave stiffness constant $D$ is obtained from the magnetic term $%
BT^{3/2}$.}
\label{tab:2}
\begin{tabular}{cccccc}
Sample & $\gamma$ (mJ/mol$\,$K$^{2}$) & $\beta$ (mJ/mol$\,$K$^{4}$) & $%
\theta_{D}$ (K) & $B$ (mJ/mol$\,$K$^{5/2}$) & $D$ (meV\AA$^{2}$) \\ \hline
1 ($\delta = 0.11$) & $23 \pm 3$ & $0.193 \pm 0.02$ & $369 \pm 13$ & $2.1
\pm 0.6$ & $75 \pm 15$ \\ 
2 ($\delta = 0.15$) & $19 \pm 2$ & $0.168 \pm 0.02$ & $387 \pm 15$ & $7.5
\pm 1.0$ & $32 \pm 3$ \\ 
3 ($\delta = 0.26$) & 0 & $0.0786 \pm 0.007$ & $498 \pm 15$ & $21.2 \pm 0.2$
& $16.1 \pm 0.1$%
\end{tabular}
\end{table}

\begin{multicols}{2}
\noindent  $\beta $ is related to the Debye temperature $\theta _{D}$, and
the coefficient $B$ to the spin-wave stiffness constant $D$.\cite{book}

The fitting parameters obtained for all samples are given in Table II, and
the fitted curves can be seen in Fig.\ 4(b) in a plot of $C$ vs.\ $T$. For
Samples 1 and 2 ($\delta $ = 0.11 and 0.15) although the plot of $C/T$ vs.\ $
T^{2}$ gives approximately straight lines, a careful fitting procedure
confirms the existence of a magnetic $BT^{3/2}$ term. The uncertainty in the
coefficients is estimated mostly by varying the fitted temperature range.
All fitted curves fall within the experimental data with a maximum
dispersion smaller than $\pm$0.7\% in more than 90\% of the points, and no
systematic departures from the fitted curves are observed. In Sample 3 ($
\delta = 0.26$) we found no contribution arising from a linear term $\gamma
T $. An upper estimate gives $\gamma < 0.8\;$ mJ/mol$\,$K$^{2}$, obtained
using a maximum fitting temperature above 9$\;$K. Below this range, the
inclusion of a linear term in the fitted expression yields negative values
of $\gamma$. By ~allowing ~the ~magnetic contribution ~to vary as ~$BT^n$

\narrowtext
\begin{figure}
\begin{center}
\includegraphics[width=8.0cm,angle=0,clip]{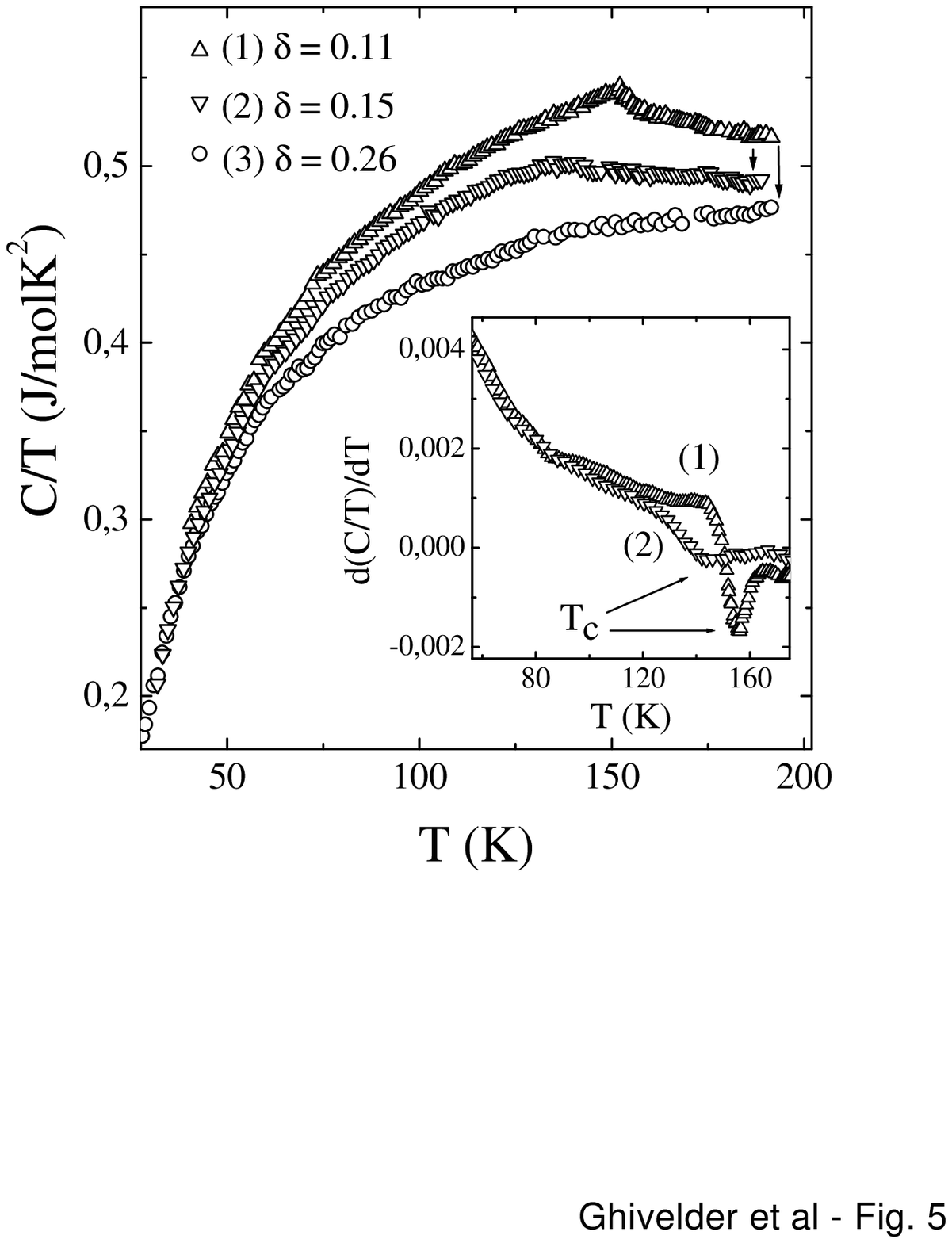}
\end{center}
\caption{High-temperature (30--200$\;$K) specific heat of LaMnO$_{3+\delta}$
, plotted as $C/T$ vs.\ $T$. Results for Samples 2 and 3 are shifted
downward for clarity, as indicated by the arrows. The inset shows the
temperature derivative $d(C/T)/dT$ vs.\ $T$ for Samples 1 and 2, with the FM
transition temperatures indicated.}
\label{fig:5}
\end{figure}

\noindent we find
a best fit with $n$ very close to the assumed value of 3/2. One of the most
important and unexpected results obtained from our low-temperature
specific-heat data is the observation of a very high linear coefficient $
\gamma$ in Samples 1 and 2. In this case, by fitting the data only with
linear and cubic terms we obtain even higher values of $\gamma$. Possible
origins of this contribution will be discussed below.

The Debye temperature $\theta_{D}$ significantly increases with the increase
of vacancy content in LaMnO$_{3+\delta}$. The values of $\theta _{D}$, in
the range 370-500$\;$K, are comparable to those previously reported in
manganite perovskites.\cite{Ghiv,Ham,Coey,Wood,Tok} The tendency to an
increase of $\theta_{D}$ with higher hole doping has been previously
observed.\cite{Ghiv,Wood,Tok} It has been argued\cite{Tok} that the
reduction of lattice stiffness at low doping values could be related to
dynamic Jahn-Teller distortion in the compounds. The large value of $
\theta_{D}$ in Sample 3, with the highest content of Mn$^{4+}$, is close to
that observed\cite{Ghiv} in the AF insulator La$_{0.38}$Ca$_{0.62}$MnO$_3$.
This suggests that AF interactions, also present in Sample 3, may contribute
to a hardening of the lattice vibrations.

The magnitude of the $BT^{3/2}$ term is also of relevance, providing
information on the spin-wave excitations in the compounds. The value of the
spin-wave stiffness constant determined for Sample 1, $D = 75 \;$meV\AA$^{2}$
, is approximately half of that obtained for La$_{0.7}$Ca$_{0.3}$MnO$_3$ ($D
= 170 \;$meV\AA$^{2}$)\cite{Lynn} and La$_{0.7}$Sr$_{0.3}$MnO$_3$ ($D = 154
\;$meV\AA$^{2}$),\cite{Smol} both in the FM metallic phase. The value of $D
= 32\;$meV\AA$^{2}$ in Sample 2 is of the same order as in the FM insulator
La$_{0.9}$Ca$_{0.1}$MnO$_3$ ($D = 40 \;$meV\AA$^{2}$),\cite{Ghiv} whose
insulating character is also interpreted as a disorder effect. This is
consistent with the fact that increasing disorder should give rise to lower
values of $D$, i.e.,``softer'' spin waves, as it is expected to reduce the
strength of the ferromagnetic coupling. The observation of a magnetic $
BT^{3/2}$ contribution in Sample 3, for which a spin-glass phase is
observed, will be addressed in the next section.

For completeness, Fig.\ 5 displays the high-temperature (30--200$\;$K)
specific heat of the investigated LaMnO$_{3+\delta }$ samples, plotted as $
C/T$ vs.\ $T$. Results for Samples 2 and 3 are shifted downward for clarity.
Sample 1 shows a small anomaly associated with the FM transition at 152$\;$
K, coinciding with the transition temperature obtained from the AC
susceptibility. No anomaly is observed in the results for Samples 2 and 3.
The inset shows the temperature derivative, $d(C/T)/dT$, for Samples 1 and
2. The FM transition in Sample 2 is visible in the derivative plot at 143$\;$
K, again coinciding with the susceptibility measurements. Phase transitions
with a large temperature width often show no specific-heat anomaly, as
reported\cite{Ghiv} in the FM insulator La$_{0.90}$Ca$_{0.10}$MnO$_3$. For
Sample 3, a specific-heat anomaly is not observed even in the derivative
plot (not shown), as expected for a spin glass. For Sample 1, the entropy
associated with the FM transition, which can be obtained from $\Delta S=\int
(C/T)dT$, is $\Delta S = 0.21 \pm 0.02\;$J/mol$\,$K. The subtracted lattice
contribution is estimated by excluding the peak region from the data, and
fitting the remaining data with a sum of three Einstein optical modes. The
value of $\Delta S$ is about an order of magnitude smaller than reported on
Ca-doped samples,\cite{Ghiv,Tanaka} and on other manganite compounds,\cite
{Nd,PrCa} where in turn the $\Delta S$ values are also smaller than expected
from the ordering of the spin system. A thorough discussion related to this
``missing'' entropy can be found elsewhere.\cite{PrCa}

\section{Discussion}

From our susceptibility and magnetization data we have established that the
FM phase of hole doped LaMnO$_{3+\delta }$ samples evolves to a
cluster-glass-like state. Several other manganite compounds present similar
behavior when substitution occurs in the manganese site. If one takes, for
instance, the standard CMR compound La$_{0.7}$Ca$_{0.3}$MnO$_{3}$,
substitution of Mn by Co\ (Ref. \onlinecite{Gay}) or In (Ref.\ 
\onlinecite{Sanchez}) also gives rise to an insulating cluster-glass phase.
This suggests that the DE interaction, mostly responsible for the metallic
FM state of doped manganites, is inhibited by random disorder in the system.
The formation of FM clusters is accompanied by strong charge-localizing
effects which yield an insulating state. Nevertheless, the size of the
clusters must be large enough for the $e_{g}$ electrons to extend over
several sites, and provide the observed FM interaction.

The most striking feature of the specific-heat data for Samples 1 and 2 is
the appearance of an unexpectedly large linear term, in excess of 19$\;$
mJ/mol$\,$K$^{2}$, although the system as whole is an insulator with respect
to transport properties. It is most important to understand the origin of
this anomalous contribution. Compared with the increasing number of
publications on doped manganite perovskite samples, relatively few reports
on heat capacity have been presented. Low-temperature data for LaMnO$_{3}$
doped with Ca,\cite{Ghiv,Ham,Coey} Sr,\cite{Coey,Wood} and Ba,\cite{Ham,Coey}
all in the metallic FM phase, observed a specific-heat linear term $\gamma $
in the range of 5--7$\;$mJ/mol$\,$K$^{2}$, associated with conduction
electrons. However, few previous investigations have reported high $\gamma $
values in insulating manganite samples: in the electron doped system La$
_{2.3}$Ca$_{0.7}$Mn$_{2}$O$_{7}$,\cite{LaY} the authors found $\gamma =41\;$
mJ/mol$\,$K$^{2}$, and in Nd$_{0.67}$Sr$_{0.33}$MnO$_{3}$\cite{Nd} a value
of $\gamma =25\;$mJ/mol$\,$K$^{2}$ was observed. A detailed explanation for
this contribution has not been put forward. As already mentioned, our
magnetic results clearly allow us to infer that ferromagnetic order in LaMnO$
_{3+\delta }$ develops in regions of limited size (clusters), whose magnetic
moments undergo a spin-glass-like transition. We will argue now that our
heat capacity results are consistent with this picture.

The stoichiometric compound LaMnO$_{3}$ ($\delta =0$) has an orthorhombic
crystal structure,\cite{ortho71} which is a distorted form of the cubic
perovskite structure. The ideal cubic system would have a FM metallic
character, with the Fermi energy lying in the middle of the $e_{g}$ band.
\cite{Sat} The splitting of the $e_{g}$ bands, due to the Jahn-Teller
distortion, leads to a small gap (1.5$\;$eV) between the Mn $e_{g}^{1}$ and $
e_{g}^{2}$ bands. This stabilizes the A-type AF order and makes the system a
Mott insulator. As we dope with holes, Mn$^{4+}$ are created, the perovskite
distortion decreases, and the Fermi level drops down into the lowest half of
the split band. Thus the system becomes metallic, with the DE mechanism
being responsible for charge transfer among Mn ions, and the consequent
polarization of the $t_{2g}$ spins that yield ferromagnetic order. However,
in non-stoichiometric LaMnO$_{3+\delta}$, the disorder introduced by random
La and Mn vacancies may cause Anderson-like localization of the electron
states close to the band edges. In contrast to what happens in the
cation-substituted compounds, the disordering effect of Mn vacancies is
strong enough for localization to be effective even at high concentrations
of Mn$^{4+}$.

Previous theoretical investigations\cite{Allub1,Allub2,Varma} confirmed that
disorder leads to charge localization in doped manganites. As is well known
since Anderson's original paper,\cite{Ander} a distribution of
site-dependent diagonal energies produces localization of the electronic
states from the edges of the bands to an energy within them which is called
the {\it mobility edge\/}. Allub and Alascio\cite{Allub1,Allub2} have shown
that, for La$_{1-x}$Sr$_x$MnO$_3$, according to the amount of disorder and
concentration of carriers, the Fermi level can cross the mobility edge to
produce a metal-insulator transition. Disorder is quantified by a
distribution width $\Gamma$. If $\Gamma$ is large, charge localization is
enhanced, and the system remains insulating, as observed in our LaMnO$
_{3+\delta}$ samples. It is worth mentioning that electron localization also
occurs in metallic manganite compounds. It has been argued\cite{Coey} that
the $e_{g}$ electrons may be localized in large wave packets due to
potential fluctuations arising from cation substitution, and additionally by
spin-dependent fluctuations due to local deviations from FM order. In LaMnO$
_{3+\delta}$ the missing Mn ions enhance these random fluctuations, favoring
charge localization.\cite{Ranno}

On the other hand, it is reasonable to assume that at low vacancy
concentration the Fermi level does not fall too far above the mobility edge,
which implies that the localization length may be fairly large. Charge
carriers can thus hop between a number of Mn ions, which actually defines
the FM clusters. The electron mobility inside the clusters ensures the
effectiveness of the DE interaction, giving rise to the observed FM
behavior. Furthermore, if the FM regions are not too small, regular spin
waves can be excited inside them, yielding the observed $T^{3/2}$
contribution to the specific heat. This is consistent with the values
obtained for the spin wave stiffness in these compounds. The electron
levels, although localized, are not largely spaced in energy, allowing for
thermal excitations that contribute with a linear term to the specific heat
as a function of temperature. It remains to be understood why the
coefficient of the specific-heat linear term is so large in our results, as
compared to other perovskite systems. A number of mass enhancement
mechanisms may be envisaged, like magnetic polarons, lattice polarons
related to the dynamical Jan-Teller effect, or Coulomb interaction effects.
However, it is not straightforward to understand why these effects would not
be equally noticeable in most doped manganite compounds. We suggest that the
explanation lies in the fact that localization has changed the Fermi level
to a region of high density of states. For instance, it is possible that the
disorder yields an enhancement of the two-dimensional character of the
bands, giving rise to a high density of states. Indeed, band structure
calculations\cite{Sat,Sing} on stoichiometric LaMnO$_3$ revealed sharp
features resembling the typical logarithmic van Hove singularities of
two-dimensional tight-binding bands.

Sample 3, with the largest vacancy content, shows qualitatively distinct
characteristics. Nevertheless, its behavior can be interpreted with the same
arguments discussed above. The localization length is now very small due to
the high degree of disorder. Thus, FM clusters are no longer formed, which
is consistent with the observed magnetic response of the compound. The
absence of a linear term in the specific heat reflects the higher degree of
localization of the charge carriers, which effectively prevents the DE
mechanism. The system behavior closely resembles that of a regular spin
glass, with short range FM interaction competing with AF coupling, the
latter arising from the high Mn$^{4+}$ content. It is somewhat puzzling,
though, that the dominant contribution to the specific heat is a term
proportional to $T^{3/2}$, which is usually attributed to spin waves in a
long range FM system. However, according to computer simulations by Walker
and Walstedt\cite{ww} on a model spin glass, the low-energy excitations are
collective modes, even though the local magnetic moments do not show long
range order. Thus, some power-law behavior of the specific heat with
temperature can be expected. Linear and quadratic terms are obtained in
Ref.\ \onlinecite{ww} for a model metallic spin glass with RKKY
interactions, but the actual value of the exponent depends on details of the
distribution of low-lying excitations, and a value of 3/2 cannot be ruled
out.

\section{Conclusions}

In this work we have presented measurements of AC susceptibility, DC
magnetization, and specific heat in a series of LaMnO$_{3+\delta}$ samples
with large $\delta$ values, and therefore high degree of disorder. The aim
is to provide a better understanding of the role of La and Mn vacancies in
the properties of mixed-valence manganites. From our analysis we may draw
two main conclusions: (i) magnetic measurements showed that the previously
known FM insulating phase of these compounds displays a disorder-induced
cluster-glass-like behavior; (ii) the anomalous high specific-heat linear
coefficient $\gamma$ in this case gives evidence of a high density of
localized states around the Fermi level, even though the latter falls in a
region of Anderson-localized states. Hence, charge localization is enhanced
due to disorder in the system, and the low temperature FM insulating state
consists of randomly oriented FM clusters, which align in small applied
fields. The carriers, though localized, may hop between several Mn sites to
ensure the DE interaction responsible for the FM order. In the sample with
the highest vacancy content the increased random disorder, and the
competition between FM and AF interactions give rise to a spin-glass state.

\section{Acknowledgments}

We thank Mucio Continentino and Gerardo Mart\'{\i }nez for helpful
discussions. This research was financed by the Brazilian Ministry of Science
and Technology under the contract PRONEX/FINEP/CNPq no 41.96.0907.00.
Additional support was also given by FUJB and FAPERJ. J.A.A. thanks the
Spanish CICyT for funds to the project PB97-1181. L.F.C. was supported by
the EPSRC grant number GR/K 73862 and by the Royal Society, U.K.

\end{multicols}

\end{document}